\documentclass{ledger}

\newcommand{\thefirstpagenum}[0]{1}
\newcommand{\thelastpagenum}[0]{5}
\newcommand{\theyear}[0]{201X}
\newcommand{\thevol}[0]{X}
\newcommand{\thedoi}[0]{DOI 10.5195/LEDGER.\theyear.X}
\newcommand{\ledgerpages}[0]{\thefirstpagenum-\thelastpagenum}

\usepackage{tikz}
\usepackage{color, colortbl}
\definecolor{Gray}{gray}{0.9}
\usepackage{svg}
\usepackage{makecell}

\hypersetup{pdfauthor={First Last}, pdftitle={This is a Title}}

\title{Exploring the Impact: How Decentralized Exchange Designs Shape Traders' Behavior on Perpetual Future Contracts}

\author{Erdong Chen\thanks{E.~Chen (ed1@e.ntu.edu.sg) is from School of Computer Science and Engineering, Nanyang Technological University, Singapore.}\and Mengzhong Ma\thanks{M.~Ma (mengzhon001@e.ntu.edu.sg) is from Interdisciplinary Graduate School, Nanyang Technological University, Singapore.}\and Zixin Nie\thanks{Z.~Nie (s230135@e.ntu.edu.sg) is from School of Computer Science and Engineering, Nanyang Technological University, Singapore.}}

\pretitle{
  \centering \selectfont RESEARCH ARTICLE \par 
  \fontsize{24pt}{28pt}\selectfont} 

\begin{document}
\setcounter{footnote}{0}
\maketitle

\thispagestyle{pagefirst}

\begin{abstract}
In this paper, we analyze traders' behavior within both centralized exchanges (CEXs) and decentralized exchanges (DEXs), focusing on the volatility of Bitcoin prices and the trading activity of investors engaged in perpetual future contracts. We categorize the architecture of perpetual future exchanges into three distinct models, each exhibiting unique patterns of trader behavior in relation to trading volume, open interest, liquidation, and leverage. Our detailed examination of DEXs, especially those utilizing the Virtual Automated Market Making (VAMM) Model, uncovers a differential impact of open interest on long versus short positions. In exchanges which operate under the Oracle Pricing Model, we find that traders primarily act as price takers, with their trading actions reflecting direct responses to price movements of the underlying assets. Furthermore, our research highlights a significant propensity among less informed traders to overreact to positive news, as demonstrated by an increase in long positions. This study contributes to the understanding of market dynamics in digital asset exchanges, offering insights into the behavioral finance for future innovation of decentralized finance.

\begin{keywords}
\item Blockchain.
\item Perpetual Futures.
\item Volatility.
\item DeFi.
\end{keywords}
\end{abstract}

\section{Introduction}

Over the past decade, blockchain technology has seen rapid development, attracting over 200 million users globally. Among its applications, cryptocurrency derivative trading, particularly perpetual futures, has become prominent, with daily trading volumes exceeding \$100 billion. Perpetual futures, introduced into the cryptocurrency market by BitMEX (\href{https://bitmex.com/}{https://bitmex.com/}) in 2016 and conceptualized by Robert J. Shiller\cite{shiller1993measuring}, represent a novel variant of traditional futures contracts. Unlike their traditional counterparts, perpetual futures do not have a fixed delivery date with their prices pegged to spot prices of the underlying assets, allowing for indefinite trading without the need for contract rollovers.Investors show a preference for perpetual futures over traditional futures due to their efficiency in hedging and speculation, high leverage, and the absence of delivery and rollover requirements. Before 2022, investors trade perpetual futures on centralized exchanges (CEXs), but following the collapse of FTX there emerged many decentralized exchanges (DEXs) carrying perpetual futures, which are operated with more elements on-chain. Since the DEXs are constructed with smart contracts involved, more transparency can be ensured, helping guard against the fraudulence incurred in CEXs.

Although both running perpetual futures, CEXs and DEXs are considerably different in terms of their order matching and price discovering, leading to different traders' behavior and implications on the market micro-structure. The design of the current CEXs and DEXs can be classified into three exchange models according to the way orders are matched: Limit-Order-Book (LOB) Model, Virtual Market Making (VAMM) Model, and Oracle Pricing Model. Unlike the case of LOB Model, which adopt the limit-order book to form prices, orders submitted by traders in exchanges of VAMM Model are matched with liquidity providers \footnote{Liquidity providers deposit assets in liquidity pools facilitated by smart contracts, where traders execute transactions.} and the prices are formed based on the relative abundance of assets in the liquidity pool. While in exchanges of LOB and VAMM Model traders' trading activity help form the price, traders in exchanges of Oracle Pricing Model accept the price offered by the Oracle and thus act as pure price takers. Although liquidity providers are still passively act as the trading counterparty, the pricing is governed by a Price Oracle, which assimilates prices from various spot exchanges, thereby rendering transactions in exchanges of Oracle Pricing Model incapable of impacting the price directly\footnote{Details about the design of different exchange models can be found in \nameref{Appendix.A} or the Systems of Knowledge paper \cite{chen2024perpetual}.}. Therefore, in this paper we ask how those differences in exchanges of difference exchange models affect price formation and investors' behavior and what kinds of empirical implications can be provided to the potential theoretical models for those exchange models.

We investigate the distinct behaviors of traders on exchanges within different models by examining the relationship between the price volatility of the underlying asset, i.e., Bitcoin, and various trading activities, i.e., trading volume, open interest, liquidation, and leverage. Our study encompasses prominent exchanges of LOB Model such as Binance (\href{https://www.binance.com/en}{https://www.binance.com/en}) and Bybit (\href{https://www.bybit.com}{https://www.bybit.com}), alongside exchanges of Oracle Pricing Model like GMX (\href{https://gmx.io}{https://gmx.io}) and GNS ({\href{https://gains.trade}{https://gains.trade}), and Perpetual Protocol V2 (\href{https://perp.com}{https://perp.com}) which employs VAMM Model. The empirical evidence shows that, for exchanges of LOB Model (Binance and Bybit), price volatility is positively related with trading volumes but negatively related with open interests, consistent with findings in the traditional future markets. While the effects of trading activity on price volatility in exchanges of LOB Model reflect the market depth and can be explained by Kyle's model \cite{kyle1985continuous}, the VAMM Model (Perpetual Protocol V2) introduces a nuanced dynamic in the impact of open interest, varying between long and short positions. This asymmetry is attributable to the VAMM's price formation mechanism, where the rate of change in an asset's relative price inversely correlates with its abundance in the liquidity pool. Consequently, market depth increases with rising open interests in short positions, as the underlying asset accumulates in the liquidity pool, and the reverse holds true for long positions. While in exchanges of LOB and VAMM Model traders' trading activity help form the price, traders in exchanges of Oracle Pricing Model (GMX and GNS) accept the price offered by the Oracle and thus act as pure price takers. Therefore, trading activity should be interpreted as purely traders' reaction to the price change of the underlying asset. When the price become more volatile, trading volumes increase while open interests decrease, with change in long and short positions different. These empirical estimations can be explained by the predictions based on Shalen’s dispersion of beliefs model (1993) \cite{shalen1993volume}, which address the asymmetry of information traders can access. We also find that uninformed traders tend to overreact more to positive news than negative, evidenced by more of the long positions accumulated.


Existing literature evaluates cryptocurrency future contracts mainly in terms of their relationship with the spot market. Akyildirim et al. \cite{akyildirim2020development} studied the impact of Bitcoin futures on the cryptocurrency market, especially the introduction of CME and CBOE futures contracts in December 2017. Alexander et al. \cite{alexander2020bitmex} found that BitMEX derivatives lead the price discovery process over Bitcoin spot markets. Hung et al. \cite{hung2021trading} identifies substantial pricing effects and breakpoints in market efficiency, indicating the dominant role of Bitcoin futures in price discovery compared to spot markets. As a special case of future contracts, perpetual future contracts are scarcely investigated, although with much higher trading volume. Besides the theoretical discussion on the arbitrage between perpetual future markets and spot markets \cite{he2022fundamentals}, there lacks enough empirical works examining perpetual future traders' behavior, especially in DEXs. After the pioneering work by Soska et al. \cite{soska2021towards}, which conducted the first analysis on the trader profile for perpetual future contracts in BitMEX, i.e., a CEX, Alexander et al. \cite{alexander2023hedging} constructed the optimal hedging strategy with empirical corroboration.

Since some DEXs of perpetual futures adopt VAMM Model, which utilize Automated Market Making (AMM) as the core, our study also shed lights on the effect of traders specifically derived by AMM. There is a emerging sort of research on the economic implication of AMM, most of which, however, focus on the incentive of liquidity provision, driven by the relationship between transaction fee and impermanent loss  \cite{milionis2022automated,hasbrouck2023economic,lehar2021decentralized,capponi2021adoption}. While the limited existing empirical studies are based on the spot market \cite{lehar2021decentralized,han2022trust}, i.e., Uniswap, our study contribute to the understanding of AMM by conducting the first analysis on the future market supported by AMM from the perspective of traders, enlightening the effect driven by AMM in a new context. 

While the primary focus of our research is on Decentralized Exchanges (DEXs) operating with perpetual future contracts, the findings and theoretical frameworks developed herein have broader applicability, extending to other trading paradigms within DEXs that exhibit the fundamental bifurcation between Traders and Liquidity Providers. This is exemplified in platforms such as PancakeSwap (\href{https://pancakeswap.finance}{https://pancakeswap.finance}) and Uniswap (\href{https://uniswap.org}{https://uniswap.org}). Furthermore, as the Decentralized Finance (DeFi) ecosystem evolves, the proliferation of synthetic assets is anticipated to enhance the diversification of portfolios managed on DEXs. The insights gleaned from this study are thus of paramount importance to both academic researchers and industry practitioners.

\section{Data and Methodology}
GMX constructs its trading system on Arbitrum and Avalanche \footnote{Arbitrum and Avalanche are two chains to address scalability of Ethereum (\href{https://vitalik.eth.limo/general/2023/10/31/l2types.html}{https://vitalik.eth.limo/general/2023/10/31/l2types.html}), while detailed explanation can be found at: \href{https://arbitrum.io/}{arbitrum.io} and \href{https://www.avax.network/}{www.avax.network}.}, and we focus on tradings on Arbitrum from 31 August 2021 to 26 September 2023, the data of which are retrieved from Dune (\href{https://dune.com/queries/3110823}{https://dune.com/queries/3110823}). Similarly, GNS is based on Arbitrum and Polygon \footnote{Information about Polygon can be found at: \href{https://polygon.technology/}{polygon.technology}.}, and we choose the perpetual futures on Polygon from 20 December 2021 to 26 September 2023 to examine, the data of which are retrieved by our constructed queries on Dune (\href{https://dune.com/queries/2829105}{https://dune.com/queries/2829105} and \href{https://dune.com/queries/2825251}{https://dune.com/queries/2825251}). We also retrieve data for Perpetual Protocol V2 (based on Etherum mainnet) from 27 November 2021 to 26 September 2023 by constructing the query on Dune (\href{https://dune.com/queries/3039245}{https://dune.com/queries/3039245}). 

We retrieve data on Binance and Bybit at the daily-level frequency using API provided by Coinalyze (\href{https://coinalyze.net/}{https://coinalyze.net/}) from 05 August 2020. Data of BTC pricing at the daily-level are collected from CryptoCompare with API (\href{https://cryptocompare.com/}{https://cryptocompare.com/}). 

It is paramount to recognize that the pricing mechanisms governing perpetual futures contracts are intentionally designed to approximate the spot price of the underlying asset. Consequently, traders engaged in the trading of perpetual futures are confronted with prices inherently tethered to the underlying asset's market valuation\cite{he2022fundamentals}. For the purpose of our analysis, and without loss of generality, we direct our focus toward an examination of how traders operating within the perpetual futures market respond to variations in the price of the underlying asset, specifically Bitcoin.

To establish an estimation for the daily volatility of Bitcoin's price, this study employs the extreme-value volatility estimator, proposed by Garman and Klass (1980) \cite{garman1980estimation}, as follows:
\begin{equation}
\begin{split}
\hat{\sigma}_t=\left\{0.5 \times\left(\ln \left(P_{t, H} / P_{t, L}\right)\right)^2\right.-
\left.(2 \ln (2)-1)\left(\ln \left(P_{t, O} / P_{t, C}\right)\right)^2\right\}^{1 / 2},\label{eq1}
\end{split}
\end{equation}
where $P_{t, H}, P_{t, L}, P_{t, O}$, and $P_{t, C}$ are the high, low, opening, and closing prices of Bitcoin on date $t$, respectively.

Previous research has consistently shown that variables related to traders' activities, such as trading volumes and open interests, exhibit significant correlation \cite{bessembinder1993price}, suggesting their predictability based on the previous values. Consequently, these variables are often analyzed by dividing them into expected and unexpected components. The expected component encompasses information about current trends, while the unexpected component captures unforeseen changes in traders' behavior. Hedgers, who adjust positions infrequently with a long-term focus, exhibit predictable behavior captured in the expected component. In contrast, speculators, who frequently adjust positions to manage risk, are represented by the unpredictable, unexpected component of activity. Following the methodology of Bessembinder and Seguin (1993) \cite{bessembinder1993price}, we partition trading volume, open interest, volume of liquidated positions, and daily average leverage using ARIMA$(p,k,q)$ model, where fitted values are regarded as the expected component and its residuals as the unexpected component.

To explore how investors' behavior vary with volatility, we adapt the model proposed by Wang and Yau (2002) \cite{wang_effect_2002} to include daily volumes of liquidated short and long positions separately:


\begin{equation}
\begin{aligned}
\hat{\sigma}_t= & \mu+\sum_{i=1}^m \phi_i \hat{\sigma}_{t-i}+\sum_{j=1}^3 \alpha_j E A_{j, t}+\sum_{j=1}^3 \beta_j U A_{j, t}+\sum_{k=1}^2 \gamma_k E L_{k, t}+\\
&\sum_{k=1}^2 \lambda_j U L_{k, t}+\varepsilon_t,\label{eq2}
\end{aligned}
\end{equation}
where $\hat{\sigma}_t$ is the estimated volatility on day $t$. $E A_{j, t}$ and $U A_{j, t}$ denote the expected and unexpected trading activity respectively, with $j$ represents different trading activity when assigned different values: $j=1$ for open interest on long positions, $j=2$ for open interest on short positions, and $j=3$ for trading volumes. $E L_{k, t}$ and $U L_{k, t}$ denote represent the expected and unexpected daily liquidated volumes for long ($k=1$) and short ($k=2$) positions. The model accounts for volatility persistence through lagged volatility estimates, with the lag structure ($m$) determined by the Akaike Information Criterion. Given the established correlation between trading activities and volatility \cite{wang_effect_2002}, these variables are included as controls. This analysis using Eq.~\eqref{eq2} is applied to perpetual futures from all three DEXs and two CEXs, given the availability of trading volume, open interest, and liquidation data.

Anticipating that traders are inclined to modify their leverage strategies in response to heightened liquidation risks under increased price volatility of the underlying asset, this study incorporates an analysis of leverage adjustments. Research focusing on centralized exchanges (CEXs) often omits leverage considerations due to the general unavailability of such data \cite{alexander_hedging_2023,soska2021towards}. However, leveraging transaction-level data from GMX and GNS, this investigation calculates and integrates the daily average leverages for both long and short positions into the analytical framework. This is achieved by adapting the existing volatility estimation model, as delineated in Equation \eqref{eq2}:
\begin{equation}
\begin{aligned}
\hat{\sigma}_t= & \mu+\sum_{i=1}^m \phi_i \hat{\sigma}_{t-i}+\sum_{j=1}^3 \alpha_j E A_{j, t}+\sum_{j=1}^3 \beta_j U A_{j, t}+\sum_{k=1}^2 \gamma_k E L_{k, t}+\\
&\sum_{j=k}^2 \lambda_j U L_{k, t}+\sum_{k=1}^2 \rho_k E LV_{k, t}+\sum_{k=1}^2 \omega_k U LV_{k, t}+\varepsilon_t,\label{eq3}
\end{aligned}
\end{equation}
where $E LV_{k, t}$ and $U LV_{k, t}$ denote expected and unexpected average leverage for long  ($k=1$) and short positions ($k=2$), respectively,  on day $t$.  Notably, due to the cross-margining feature in Perpetual Protocol v2, where a collective pool of funds backs each position \footnote{Documents of Leverage and Margin Ratio for Perpetual Protocol v2: \href{https://support.perp.com/hc/en-us/articles/5257393633945-Leverage-Margin-Ratio}{support.perp.com/hc/en-us/articles/5257393633945-Leverage-Margin-Ratio}.}, the specific leverage level for each position remains undefined. Consequently, the leverage analysis using Equation \eqref{eq3} is confined to perpetual futures from GMX and GNS.

\section{Empirical Results}

\subsection{Empirical results and analysis for LOB Model (Binance and Bybit)}

\begin{table}
\caption{Regression Results on Eq.~\eqref{eq2}}\label{tab_eq2}
\begin{tabularx}{\textwidth}{ c  c  c  C  C  C }
\hline
\tableheading{Variables} &   \multicolumn{5}{c}{\tableheading{Exchanges with Perpetual Contracts on Bitcoin}\footnote{This table reports the regression results of Eq.~\eqref{eq2}.}\footnote{For perpetual futures on both DEXs and CEXs, contract size is undefined and traders trade perpetual futures with arbitrary units of Bitcoins, so trading volumes and open interests are measured in USD.}}  \\ 
\cline{2-6}
 & \fontsize{8pt}{8pt}\selectfont Binance & \fontsize{8pt}{8pt}\selectfont Bybit\footnote{Since for CEXs each future contract is matched with a buyer and a seller, resulting in the same dollar value of open interests for long and short, we regress price volatility only on open interest on short.} & \fontsize{8pt}{8pt}\selectfont Perpetual Protocol V2 & \fontsize{8pt}{8pt}\selectfont GMX & \fontsize{8pt}{8pt}\selectfont GNS  \\
 \hline
Intercept\footnote{In each cell, the t-statistics are in the parentheses. $*$, $**$, and $***$ denote significance at 0.1, 0.05, and 0.01 level, respectively.} & 
\makecell{0.011\\(6.27)***} & 
\makecell{2.145\\(0.15)}&
\makecell{0.009\\(3.48)***} &
\makecell{1.438e-12\\(16.82)***} & 
\makecell{0.017\\(8.82)***}
\\
Lagged volatility & 
\makecell{0.200\\ (10.16)***} & 
\makecell{0.210\\ (17.19)***}&
\makecell{0.252\\ (8.19)***} &
\makecell{0.256\\ (8.10)***} & 
\makecell{0.371\\ (10.57)***}
\\
\hline
\multicolumn{6}{l}{Trading Activity:}
\\
\makecell{Expected trading\\volume} &
\makecell{3.06e-08\\ (12.87)***} & 
\makecell{6.041e-08\\ (8.91)***}&
\makecell{2.59e-09\\ (8.76)***} &
\makecell{1.186e-10\\(6.30)***} & 
\makecell{1.222e-11\\ (0.19)} 
\\
\makecell{Unexpected trading\\volume}& 
\makecell{5.943e-08\\(24.07)***} & 
\makecell{1.409e-07\\(15.62)***}&
\makecell{3.62e-09\\(14.42)***} & 
\makecell{1.29e-10\\(9.83)***} & 
\makecell{4.53e-10\\(6.65)***} 
\\
Expected OI on short & 
\makecell{-1.11e-12\\(-2.24)***} & 
\makecell{-8.55e-12\\(-8.69)***}&
\makecell{-1.678e-09\\(-4.40)***} &
\makecell{-4.765e-10\\(-7.36)***} & 
\makecell{-2.059e-10\\(-1.99)**} 
\\
Unexpected OI on short & 
\makecell{-2.365e-11\\(-8.93)***} & 
\makecell{-3.138e-11\\(-4.65)*** } &
\makecell{1.125e-08\\(4.53)*** } &
\makecell{-2.808e-10\\(-3.27)*** } & 
\makecell{-4.654e-10\\(-3.08)*** } 
\\
Expected OI on long & 
- & 
- &
\makecell{3.569e-09\\(3.98)***} &
\makecell{-1.88e-10\\(-5.99)***} & 
\makecell{2.98e-11\\(0.44)}
\\
Unexpected OI on long & 
- & 
- &
\makecell{1.481e-09\\(0.161)} &
\makecell{-1.75e-10\\(-2.16)**} & 
\makecell{-4.261e-11\\(-0.29)}
\\
\hline
\multicolumn{6}{l}{Liquidation:}
\\
\makecell{Expected liquidated\\volume on short positions} &
\makecell{2.864e-10\\(5.46)***} & 
\makecell{-2.832e-10\\(-2.32)**}&
\makecell{-5.983e-07\\(-2.09)**} &
\makecell{2.449e-09\\(1.09)} & 
\makecell{-1.483e-09\\(-1.26)}
\\
\makecell{Unexpected liquidated\\volume on short positions} &
\makecell{1.655e-10\\(6.95)***} & 
\makecell{5.098e-10\\(7.22)***}&
\makecell{-3.032e-08\\(-1.61)} &
\makecell{2.007e-09\\(1.61)}  & 
\makecell{3.386e-10\\(0.835)}
\\
\makecell{Expected liquidated\\volume on long positions}& 
\makecell{8.534e-11\\(3.00)***} & 
\makecell{1.672e-09\\(11.96)**} &
\makecell{5.008e-07\\(4.16)***} & 
\makecell{1.82e-07\\(17.71)***} & 
\makecell{7.295e-09\\(3.26)***}
\\
\makecell{Unexpected liquidated\\volume on long positions} &
\makecell{7.143e-11\\(6.82)***} & 
\makecell{1.278e-09\\(16.62)***} &
\makecell{1.482e-07\\(5.39)***} &
\makecell{2.593e-09\\(5.08)***} & 
\makecell{1.448e-09\\(1.46)} 
\\
\hline
Adjusted $R^{2}$ & 
0.600 & 
0.600 &
0.466 &
0.380 & 
0.282  
\\
AIC & 
-6435 & 
-6511 &
-4018 &
-4389 & 
-3649
\\
No. of obs. & 
1147 & 
1147 &
668 &
747 & 
628 \\
\hline

\end{tabularx}
\end{table}


\cref{tab_eq2} aggregates regression results on Eq.~\eqref{eq2}. Coefficient estimates for expected and unexpected trading volume are positive across all the five exchanges, consistent with the findings in the previous studies on traditional futures market \cite{bessembinder1993price,wang_effect_2002}. 

For exchanges of LOB Model, i.e., Binance and Bybit,  all the estimated coefficients measuring the marginal effects of expected and unexpected open interest are negative and significant, while the estimated coefficients for the unexpected portion are also larger in magnitude than those for the expected portion. Since in these exchanges each seller is matched with a buyer of the perpetual futures, aggregated value of long positions equals to that of short positions, so do the open interests. We only include open interests on short in Eq.~\eqref{eq2}, as open interests on long are of the same value. Since exchanges of LOB Model adopt the limit-order-books, it relates to the determinants of market depth, which Kyle (1985)\cite{kyle1985continuous} defines as the volume of unanticipated order flow required to move price by one unit. When capital inflows into the market, the liquidity increases, which is reflected by higher level of open interests, making the price less volatile in response to new orders. Consequently, it is expected that when open interest is large the price volatility conditional on contemporaneous trading volume, which proxies market depth, would be lower. Therefore, those negative estimated coefficients imply that an increase in unexpected open interest, e.g., inflow of speculative capital, lessens the impact of trading volume shock (unexpected volume) on price volatility. In our case, for Binance, when a trader increase his/her position by 1 USD, the marginal effect of the 1 USD increase in trading volume on  price volatility is $5.943\times10^{-8}$, while mitigated by the effect of 1 USD increase in open interest ($-2.365\times10^{-11}$), resulting in a smaller aggregated effect ($5.943\times10^{-8}-2.365\times10^{-11}$). On the one hand, estimated coefficients for trading volumes and open interests have opposite signs, while on the other hand, whether the marginal effect of volume on price volatility is enlarged or mitigated depends on whether open interest is reduced or increased. Trades closing, decreasing, or liquidating positions actually reduce open interest, enlarging the effect of trading volume on price volatility \footnote{Trades opening positions or rising positions increases open interests.}. Price would move with larger distance, as less liquidity reacts to absorb a trade. This is reflected by that seven out of eight coefficients for liquidated volume in Binance and Bybit are significant and positive, with expected liquidated volume on short positions for Bybit as the exception. Thus, all else being equal, an unexpected rise in liquidation reduces market depth, via lowering open interest and liquidity.


\subsection{Empirical results and analysis for VAMM Model (Perpetual Protocol V2).}

Different from the case of LOB Model, where traders are matched by the market maker and provide liquidity, traders in exchanges of VAMM Model are matched with liquidity pools, in which liquidity providers act passively as the counterparty. Traders' open interest thus does not necessarily influence the market depth in the same way as in LOB Model and the theory of market depth by Kyle (1985)\cite{kyle1985continuous} for traditional futures market is inapplicable in this context. For Perpetual Protocol V2, which adopts VAMM as the pricing mechanism, liquidity providers deposit stable coins in the vault based on which the clearing house mints virtual tokens and places those virtual tokens into the Uniswap v3 AMM \footnote{Details of how VAMM works in Perpetual Protocol V2 can be found in \href{https://support.perp.com/hc/en-us/articles/9594157347481-How-It-Works}{https://support.perp.com/hc/en-us/articles/9594157347481-How-It-Works}.}. For illustration, assume the price of the Bitcoin perpetual is determined by the exchange function:
\begin{equation}\label{eq.xyk}
\begin{aligned}
Q_{vUSDC}\times Q_{vBTC}=k
\end{aligned}
\end{equation}
where $Q_{vUSDC}$ and $Q_{vBTC}$ denote the quantities of virtual USDC (vUSDC) and virtual Bitcoin (vBTC) in the liquidity pool, and  $k$ reflects the depth of liquidity, which is determined by liquidity providers. With constant $k$, the price of virtual Bitcoin is determined by the ratio of $Q_{vUSDC}$ and $Q_{vBTC}$:
\begin{equation}\label{eq.xyk_pricing}
\begin{aligned}
P_{vBTC}=\frac{Q_{vUSDC}}{Q_{vBTC}}=k \times Q^{-2}_{vBTC}.
\end{aligned}
\end{equation}
The absolute value of the first derivative of $P_{vBTC}$ with respect to $Q_{vBTC}$ is monotonously decreasing, which can be derived as:
\begin{equation}\label{eq.xyk_pricng_change_rate}
\begin{aligned}
\left|\frac{\partial P_{v B T C}}{\partial Q_{v B T C}}\right|=\left|-2 \times k \times Q^{-3}_{vBTC}\right|=2 \times k \times Q^{-3}_{vBTC},
\end{aligned}
\end{equation}
indicating the decreasing rate of change in pricing as vBTC goes abundant in the liquidity pool. The more vBTC there are in the liquidity pool, the lower volatile its price is. When a trader opens a long position, the clearing house swaps vUSDC for vBTC from the liquidity pool, adding vUSDC to and withdrawing vBTC from the pool. $Q_{vBTC}$ decreases while $P_{vBTC}$ becomes more volatile. 

Besides, Uniswap v3 AMM implements concentrated liquidity to enhance capital efficiency, where liquidity providers specify a price range, $[\underline{P}, \bar{P}]$, within which to add liquidity \footnote{For details of geometric explanation on concentrated liquidity, readers can refer to \cite{mohan2022automated}.}. If there is only a single liquidity provider specifying the price interval $\left[P_{vBTC}^A, P_{vBTC}^B\right]$ to add liquidity, where $P_{vBTC}^A=\frac{Q_{vUSDC}^A}{Q_{vBTC}^A}$ and $P_{vBTC}^B=\frac{Q_{vUSDC}^B}{Q_{vBTC}^B}$, this concentrated liquidity can be characterized by drawing a new set of axes, i.e., $\widehat{Q_{v B T C}}$ and $\widehat{Q_{vUSDC}}$ with the origin $\hat{O}$ as shown in Fig.~\ref{fig.xyk}, to track the real reserve offered in range $AB$. While the exchange function is defined in the $Q_{v B T C}-Q_{vUSDC}$ space ($virtual$ reserve termed by Unisawp v3), the actual reserve specified in the price range is called $real$. Any point $(Q_{vBTC},Q_{vUSDC})$ in the $virtual$ space can be transformed to a point $(\widehat{Q_{vBTC}},\widehat{Q_{vUSDC}})=(Q_{vBTC}-Q_{vBTC}^B,Q_{vUSDC}-Q_{vUSDC}^A)$ in the $real$ space. The liquidity provider actually defines the coordinate of the $real$ origin $\hat{O}$, which is a function of the interval $\left[P_{vBTC}^A, P_{vBTC}^B\right]$.

\tikzset{every picture/.style={line width=0.75pt}} 
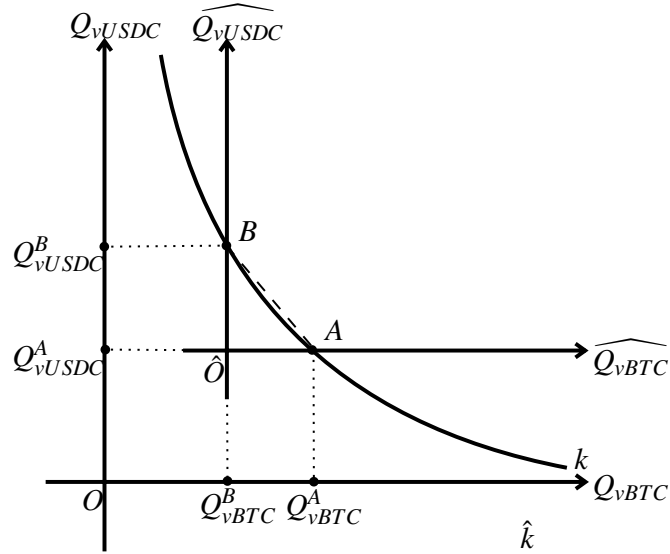
\begin{figure}
\centering
\begin{tikzpicture}[x=0.75pt,y=0.75pt,yscale=-0.7,xscale=0.7]

\draw [line width=1.5]  (111,359.91) -- (500,359.91)(153.4,42) -- (153.4,410) (493,354.91) -- (500,359.91) -- (493,364.91) (148.4,49) -- (153.4,42) -- (158.4,49)  ;
\draw  [line width=1.5]  (194,52) .. controls (222.87,218.69) and (320.39,317.93) .. (486.55,349.7) ;
\draw  [fill={rgb, 255:red, 0; green, 0; blue, 0 }  ,fill opacity=1 ] (301.5,359.75) .. controls (301.5,358.23) and (302.73,357) .. (304.25,357) .. controls (305.77,357) and (307,358.23) .. (307,359.75) .. controls (307,361.27) and (305.77,362.5) .. (304.25,362.5) .. controls (302.73,362.5) and (301.5,361.27) .. (301.5,359.75) -- cycle ;
\draw  [fill={rgb, 255:red, 0; green, 0; blue, 0 }  ,fill opacity=1 ] (151.25,264.5) .. controls (151.25,262.98) and (152.48,261.75) .. (154,261.75) .. controls (155.52,261.75) and (156.75,262.98) .. (156.75,264.5) .. controls (156.75,266.02) and (155.52,267.25) .. (154,267.25) .. controls (152.48,267.25) and (151.25,266.02) .. (151.25,264.5) -- cycle ;
\draw [line width=0.75]  [dash pattern={on 0.84pt off 2.51pt}]  (304,265.5) -- (304.25,357.75) ;
\draw [line width=0.75]  [dash pattern={on 0.84pt off 2.51pt}]  (158,264.5) -- (304,265.5) ;
\draw  [fill={rgb, 255:red, 0; green, 0; blue, 0 }  ,fill opacity=1 ] (238.5,189.5) .. controls (238.5,187.98) and (239.73,186.75) .. (241.25,186.75) .. controls (242.77,186.75) and (244,187.98) .. (244,189.5) .. controls (244,191.02) and (242.77,192.25) .. (241.25,192.25) .. controls (239.73,192.25) and (238.5,191.02) .. (238.5,189.5) -- cycle ;
\draw  [fill={rgb, 255:red, 0; green, 0; blue, 0 }  ,fill opacity=1 ] (300.25,264.75) .. controls (300.25,263.23) and (301.48,262) .. (303,262) .. controls (304.52,262) and (305.75,263.23) .. (305.75,264.75) .. controls (305.75,266.27) and (304.52,267.5) .. (303,267.5) .. controls (301.48,267.5) and (300.25,266.27) .. (300.25,264.75) -- cycle ;
\draw [line width=0.75]  [dash pattern={on 0.84pt off 2.51pt}]  (158.25,190.25) -- (241.25,189.5) ;
\draw [line width=0.75]  [dash pattern={on 0.84pt off 2.51pt}]  (241.25,189.5) -- (242,356.5) ;
\draw  [fill={rgb, 255:red, 0; green, 0; blue, 0 }  ,fill opacity=1 ] (239.25,359.25) .. controls (239.25,357.73) and (240.48,356.5) .. (242,356.5) .. controls (243.52,356.5) and (244.75,357.73) .. (244.75,359.25) .. controls (244.75,360.77) and (243.52,362) .. (242,362) .. controls (240.48,362) and (239.25,360.77) .. (239.25,359.25) -- cycle ;
\draw  [dash pattern={on 4.5pt off 4.5pt}]  (241.25,189.5) -- (305,264.75) ;
\draw  [fill={rgb, 255:red, 0; green, 0; blue, 0 }  ,fill opacity=1 ] (150.5,190.25) .. controls (150.5,188.73) and (151.73,187.5) .. (153.25,187.5) .. controls (154.77,187.5) and (156,188.73) .. (156,190.25) .. controls (156,191.77) and (154.77,193) .. (153.25,193) .. controls (151.73,193) and (150.5,191.77) .. (150.5,190.25) -- cycle ;
\draw [line width=1.5]  (210,265.17) -- (499.75,265.17)(241.58,42.29) -- (241.58,300.29) (492.75,260.17) -- (499.75,265.17) -- (492.75,270.17) (236.58,49.29) -- (241.58,42.29) -- (246.58,49.29)  ;

\draw (503,352.4) node [anchor=north west][inner sep=0.75pt]    {$Q_{vBTC}$};
\draw (126,19.4) node [anchor=north west][inner sep=0.75pt]    {$Q_{vUSDC}$};
\draw (490,333.4) node [anchor=north west][inner sep=0.75pt]    {$k$};
\draw (451,385.4) node [anchor=north west][inner sep=0.75pt]    {$\hat{k}$};
\draw (310.6,240.6) node [anchor=north west][inner sep=0.75pt]    {$A$};
\draw (248,169.8) node [anchor=north west][inner sep=0.75pt]    {$B$};
\draw (85,180.9) node [anchor=north west][inner sep=0.75pt]    {$Q_{vUSDC}^{B}$};
\draw (85,255.4) node [anchor=north west][inner sep=0.75pt]    {$Q_{vUSDC}^{A}$};
\draw (220,362.4) node [anchor=north west][inner sep=0.75pt]    {$Q_{vBTC}^{B}$};
\draw (282,362.9) node [anchor=north west][inner sep=0.75pt]    {$Q_{vBTC}^{A}$};
\draw (135,363.4) node [anchor=north west][inner sep=0.75pt]    {$O$};
\draw (214,13.4) node [anchor=north west][inner sep=0.75pt]    {$\widehat{Q_{vUSDC}}$};
\draw (503,254.4) node [anchor=north west][inner sep=0.75pt]    {$\widehat{Q_{vBTC}}$};
\draw (222,266.4) node [anchor=north west][inner sep=0.75pt]    {$\hat{O}$};

\end{tikzpicture}
\caption{A Uniswap v3 pool with a single liquidity provider.}\label{fig.xyk}
\end{figure}

In practice, liquidity offered by multiple liquidity providers cumulates around the current price level, as illustrated in Fig.~\ref{fig.liquidity_distribution}. Trades at the tails of the distribution lead to larger price move, i.e., thus larger price volatility, due to lower liquidity than at the center. Therefore, while the design of concentrated liquidity determines market depth locally, the exchange function Eq.~\eqref{eq.xyk} forms price globally. In the long-run, $(Q_{vBTC},Q_{vUSDC})$ moves along the exchange function and the change in slope determines the price volatility. In the short-run, the liquidity distribution centered at the current average price determines price volatility.

\begin{figure}[htb]
\centerline{\includegraphics[width=1\linewidth]{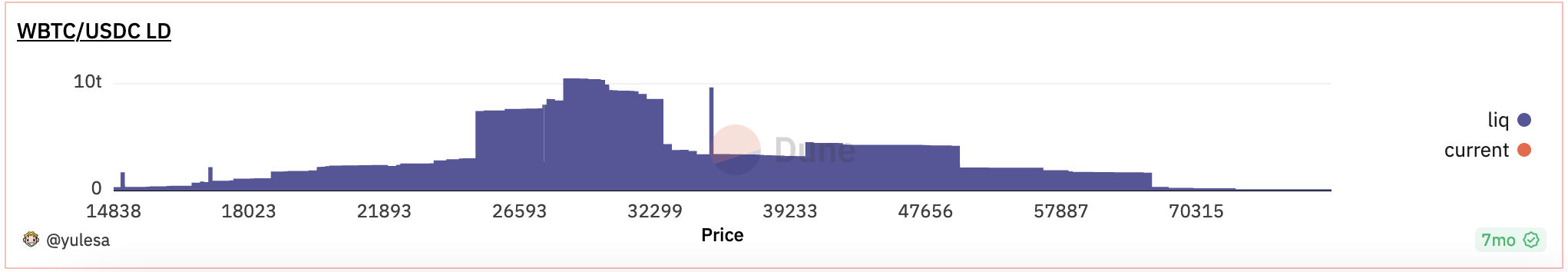}}
\caption{Example of Liquidity Distribution on Uniswap v3 AMM. This figure shows the liquidity distribution for the pool of the pair WBTC/USDC on Uniswap V3. This snapshot is made on 7 Nov 2023. Source: \href{https://dune.com/queries/65034/129883}{https://dune.com/queries/65034/129883}.}
\label{fig.liquidity_distribution}
\end{figure}

The implication of Uniswap v3 AMM on pricing volatility is consistent with our empirical findings. In Table~\ref{tab_eq2}, the coefficient for expected open interest on short is estimated to be significantly negative ($-1.678\times 10^{-9}$) while that on long is estimated to be significantly positive ($3.569\times 10^{-9}$), indicating a negative association between expected open interest on short and price volatility while a positive association for long. Since the expected portion of open interest reflects traders' long-run behavior, e.g., hedgers, in cope with the trend, the change could be mainly be described by Eq.~\eqref{eq.xyk}. When the expected open interest on short rises, the clearing house retrieve vBTC from traders' account and credit to the balance of the liquidity pool. The increased vBTC balance in the liquidity pool helps decrease the rate of pricing change and thus reducing price volatility according to Eq.~\eqref{eq.xyk_pricng_change_rate}. Besides, when expected open interest on long increases, the clearing house withdraw vBTC from the pool and credit to traders', decreasing vBTC balance in the liquidity pool and thus increasing price volatility.

While the expected portion of open interest captures how market status moves along the exchange function specified by Eq.~\eqref{eq.xyk}, the unexpected portion reflects traders' unanticipated behavior, e.g., speculation, on the basis of liquidity distribution. In the short-run, liquidity providers are lagged to adjust their bound of price within which to add liquidity, i.e., $[\underline{P}, \bar{P}]$, especially when facing unanticipated change. Therefore, open interest on both long and short positions would drive price to the tails of the liquidity distribution, enlarging pricing volatility due to lower liquidity. In Table~\ref{tab_eq2}, the estimated coefficients for the association between price volatility and unexpected open interest are all positive, i.e., $1.125\times 10^{-8}$ for short and $1.481\times 10^{-9}$ for long, substantiating our prediction based on the Uniswap v3 AMM pricing mechanism.

In the context of Perpetual Protocol V2, liquidity providers have the capability to contribute liquidity utilizing leverage, thereby incurring a risk of passive liquidation when counterparties are liquidated. Transactions that liquidate liquidity providers' positions during a price increase are categorized as \textit{liquidation on short}, whereas those during a price decline are classified as \textit{liquidation on long}. The liquidation of liquidity providers' positions results in a reduction of market depth, which is hypothesized to have a direct correlation with increased price volatility and the frequency of liquidations among liquidity providers.

Regarding the expected liquidation volume of traders' positions, its impact on price volatility is anticipated to be inversely proportional to the effect of increasing expected open interest. Specifically, an expected increase in liquidation on short positions, which reduces open interest in these positions, is posited to have a positive correlation with price volatility, and the converse holds true for long positions. In the case of the unexpected component of liquidation, its influence on price volatility aligns with that of unexpected open interest, with all unforeseen trading activities driving price movements towards the extremities of the liquidity distribution, thereby exacerbating price volatility.

Overall, an increase in unexpected liquidation in both long and short positions is expected to elevate price volatility. This is empirically corroborated in Table~\ref{tab_eq2}, where the estimated coefficient for unexpected liquidated volume on long positions is significantly positive ($5.008\times 10^{-7}$). Conversely, the aggregated effect of an increase in expected liquidation on short positions should theoretically be positive. However, this is contradicted by the negative sign of the estimated coefficient for expected liquidated volume on short positions ($-5.983\times 10^{-7}$). A plausible explanation for this anomaly is that a price rise, typically accompanying an increase in short position liquidations, may attract additional liquidity providers. This influx of liquidity providers potentially mitigates price volatility, with their dampening effect on volatility surpassing the amplifying impact of liquidations. We have conducted the Granger Causality Test to examine the relationship between between the return on Bitcoin and the net change in liquidity in the vBTC-vUSDC pool on Perpetual Protocol V2, which you can find in \nameref{Appendix.B}. It is shown that return on Bitcoin Granger-cause the net change in liquidity in the vBTC-vUSDC pool, while the reversed relationship does not hold.

The asymmetrical effect of open interest on long and short positions is also predicted in the theoretical work by \cite{aoyagi2021coexisting}, which suggests that consuming liquidity involves a larger price impact than adding it, resulting in the larger marginal cost of a ETH buy order than the marginal benefit of a ETH sell order. The model proposed in \cite{aoyagi2021coexisting} implies an ambiguous reaction of AMM liquidity to asset price volatility, driven by the migration of both traders and liquidity providers between CEXs and DEXs. In our test, since liquidation explicitly includes behavior of both traders and liquidity providers, the ambiguity mentioned in \cite{aoyagi2021coexisting} may help explain the inconsistency between our predicted effect of expected liquidation on short positions and the estimated coefficient.

\subsection{Empirical results and analysis for Oracle Pricing Model (GMX and GNS).}

In exchanges of Oracle Pricing Model such as GMX and GNS, traders accept prices as a given, lacking the capacity to exert direct influence on price movements. The pricing mechanism in these DEXs is governed by a Price Oracle, which assimilates prices from various exchanges, thereby rendering transactions on GMX and GNS incapable of impacting the price directly\footnote{The code for GMX's price feeding smart contract is publicly accessible at: \href{https://github.com/gmx-io/gmx-contracts/blob/master/contracts/oracle/FastPriceFeed.sol}{https://github.com/gmx-io/gmx-contracts/blob/master/contracts/oracle/FastPriceFeed.sol}.}. As a result, trading activities on these platforms, encompassing trading volume and open interest, merely reflect traders' interpretations and reactions to the information encapsulated within the Price Oracle's determined prices.

Drawing upon Shalen’s dispersion of beliefs model (1993) \cite{shalen1993volume}, which posits a market dichotomy between asymmetrically informed traders, market participants can be categorized into informed and uninformed traders. Uninformed traders, lacking access to private information, often engage in irrational trading based on market noise and tend to overreact to information. They strive, albeit unsuccessfully, to discern private information and market trends from current price fluctuations, leading to heightened position-taking during periods of increased price volatility. In contrast, informed traders, operating on the basis of their private information, exhibit relatively consistent beliefs over time and typically engage in trading within a constrained price range, resulting in a negative correlation between their positions and price volatility \cite{wang_effect_2002}.

In Table~\ref{tab_eq2}, the estimated coefficients that measure the relationship between open interest and price volatility are predominantly negative, except the case of long open interests in GNS. This indicates a more pronounced influence of informed traders' behavior, leading to the negative signs in these estimations. The impact of uninformed traders on open interest is effectively counterbalanced by the actions of informed traders. Notably, traders' behaviors exhibit asymmetry between buyers and sellers, as evidenced by less negative (in the case of GMX) or even insignificant (for GNS) coefficients for long positions. This asymmetry could be attributed to uninformed traders' tendency to overreact more to positive news than negative, leading to an increased propensity for long position accumulation\footnote{Existing literature on traditional spot and future markets provides evidence of traders' asymmetric reactions to positive and negative news \cite{mcqueen1996delayed,wang_effect_2002}.}. This asymmetry is further evidenced by variations in liquidated volume in relation to price volatility. As indicated in Table~\ref{tab_eq2}, for GMX and GNS, the coefficients linking liquidated volume on long positions are significantly positive, whereas those for short positions are not statistically significant. 

Aligned with Shalen’s model, it is anticipated that uninformed traders are more likely to increase leverage during periods of heightened price volatility to maximize capital efficiency, thereby enlarging their positions at a lower cost. In contrast, informed traders are less inclined to risk capital under volatile conditions, leading to a negative correlation between leverage and price volatility. Table~\ref{tab_eq3} delineates the estimated results for Eq.\eqref{eq3}. All eight coefficients relating price volatility and average leverage are estimated to be negative, indicating a more pronounced change in leverage as influenced by informed traders compared to uninformed traders. 

\begin{table}[ht]
\centering
\caption{Regression results on Eq.~\eqref{eq3}}\label{tab_eq3}
\begin{tabularx}{\textwidth}{ c  c  c c c c }
\hline
 &   \multicolumn{5}{c}{\tableheading{Exchanges with Perpetual Contracts on Bitcoin}\footnote{This table reports the regression results of Eq.~\eqref{eq3}.}\footnote{For perpetual futures on both DEXs and CEXs, contract size is undefined and traders trade perpetual futures with arbitrary units of Bitcoins, so trading volumes and open interests are measured in USD.}}  \\ 
\cline{2-3}
\cline{5-6}
\tableheading{Variables} & \fontsize{8pt}{8pt}\selectfont GMX & \fontsize{8pt}{8pt}\selectfont GNS  & \tableheading{Variables} & \fontsize{8pt}{8pt}\selectfont GMX & \fontsize{8pt}{8pt}\selectfont GNS \\
 \hline
Intercept\footnote{In each cell, the t-statistics are in the parentheses. $*$, $**$, and $***$ denote significance at 0.1, 0.05, and 0.01 level, respectively.} & 
\makecell{2.287e-12\\(9.47)***} &  
\makecell{0.028\\(3.84)***}&
Lagged volatility & 
\makecell{0.229\\ (7.26)***} & 
\makecell{0.369\\ (10.66)***} 
\\
\hline
 &\multicolumn{2}{c}{Expected portion} & &\multicolumn{2}{c}{Unexpected portion} \\

\multicolumn{3}{l}{Trading Activity:}
\\
Trading volume & 
\makecell{1.051e-10\\ (5.60)***} & 
\makecell{5.201e-11\\ (0.83)}&
 & 
\makecell{1.359e-10\\(10.41)***} & 
\makecell{4.593e-10\\(6.86)***}
\\
OI on short & 
\makecell{-4.405e-10\\(-6.52)***} & 
\makecell{-1.626e-10\\(-1.58)} &
 & 
\makecell{-2.485e-10\\(-2.89)*** } & 
\makecell{-4.423e-10\\(-2.98)*** } 
\\
OI on long & 
\makecell{-1.373e-10\\(-4.11)***} & 
\makecell{3.964e-11\\(0.58)} &
 &
\makecell{-1.091e-10\\(-1.36)} & 
\makecell{-5.505e-11\\(-0.38)}
\\
\hline
\multicolumn{3}{l}{Liquidation:}
\\
\makecell{Liquidated volume\\ on short positions} & 
            \makecell{2.538e-09\\(1.15)} & 
            \makecell{-1.467e-09\\(-1.27)} &
 &
            \makecell{1.891e-09\\(1.55)} & 
            \makecell{3.177e-10\\(0.80)} 
\\
\makecell{Liquidated volume\\ on long positions} & 
            \makecell{2.77e-07\\(9.14)***} & 
            \makecell{7.573e-09\\(3.45)***} &
 & 
            \makecell{2.539e-09\\(5.06)***} & 
            \makecell{1.384e-09\\(1.42)}
            \\
\hline
\multicolumn{3}{l}{Leverage (average level):}
\\
\makecell{Average leverage\\on short positions} & 
            \makecell{-0.0004\\(-2.21)**} & 
            \makecell{-0.0001\\(-1.40)} &
 & 
            \makecell{-0.0002\\(-3.00)***} & 
            \makecell{-0.0002\\(-4.37)***}
            \\
\makecell{Average leverage\\on long positions} & 
            \makecell{-0.0004\\(-2.15)**} & 
            \makecell{-1.193e-05\\(-0.36)} &
             & 
            \makecell{-0.0003\\(-3.65)***} & 
            \makecell{-7.237e-05\\(-1.93)**}
            \\

\hline

Adjusted $R^{2}$ & 0.398 & 0.339 &  AIC& -4405 &-3693
\\
No. of obs. & 747 &628 & & & \\
\hline

\end{tabularx}
\end{table}


In summary, the analysis reveals that informed traders exhibit a higher level of activity compared to their uninformed counterparts on GMX and GNS, a conclusion substantiated by empirical data pertaining to the dynamics of open interests and leverages. Additionally, the observed asymmetry in open interests and liquidations between long and short positions suggests a tendency among traders to disproportionately overreact to good news. However, it is imperative to acknowledge a critical distinction for analytical purposes: unlike GMX and GNS, centralized exchanges (CEXs) possess a price discovery function, leading to a reciprocal influence between traders' behavior and market prices. Under Efficient-Market Hypothesis \cite{fama1970efficient}, the price aggregate and reflect all the current private information and information contained in the past pricing history. Consequently, it is methodologically unsound to interpret changes in trading activities, e.g., volume, open interest, liquidation, and leverage, solely as reactions to price fluctuations. This perspective challenges the conclusions drawn by \cite{soska2021towards}, who posited that changes in liquidation are merely passive responses to price movements, thereby overlooking the role of liquidation as a contributory factor in price formation within BitMEX, a CEX. This oversight underscores the necessity for a nuanced understanding of the interplay between trader behavior and price dynamics, particularly in the context of CEXs where the price discovery process is inherently more complex.

\section{Conclusion}

This study underscores the necessity of developing new theoretical models, moving beyond conventional frameworks like Kyle's model\cite{kyle1985continuous}, to aptly comprehend the intricacies of DEXs. Our empirical results  uncover distinct variations in DEXs, particularly under the VAMM Model, where open interest impacts long and short positions differently. The behavior of traders in exchanges of Oracle Pricing Model (GMX and GNS), who act as price takers, is found to be consistent with Shalen’s dispersion of beliefs model\cite{shalen1993volume}. Our study highlights the significant fundamental difference between CEXs and DEXs, based on which theoretical frameworks specifically for DEX design can be derived in the future works. Table~\ref{tab:summary_empirical_results} encapsulates a summary of the unique trading behaviors elicited by the varied designs of these models. Any exchange classified into one of our three exchange models is believed to derive distinctive behavior of the traders caused by the design of the exchange model.

\begin{table}[ht]
\centering
\caption{Summary of Empirical Results}\label{tab:summary_empirical_results}
\begin{tabularx}{\textwidth}{ c  c  c c c c }
\hline
\tableheading{Exchange Model}\footnote{This table summarizes our empirical results regarding traders' behavior in exchanges of different exchange models.} &  
\tableheading{Price\\ Discovering} & 
\tableheading{Price Volatility \& \\Trading Volume} &  
\tableheading{Price Volatility \& \\ Open Interest} &
\tableheading{Symmetry; \\ Explanation}\footnote{In this table, we care about the symmetry of effects between buyers and sellers.} &
\tableheading{Exchange \\Analyzed}
\\ 
\fontsize{8pt}{8pt}\selectfont LOB Model&    
         \fontsize{8pt}{8pt}\selectfont Yes &  
         \fontsize{8pt}{8pt}\selectfont \makecell{Positively \\ correlated} &
         \fontsize{8pt}{8pt}\selectfont \makecell{Negatively \\ correlated} &
         \fontsize{8pt}{8pt}\selectfont \makecell{Symmetrical; \\Kyle's model \cite{kyle1985continuous}} &
         \fontsize{8pt}{8pt}\selectfont Binance; Bybit\\ 
         \hline 
         
         \fontsize{8pt}{8pt}\selectfont VAMM Model &    
         \fontsize{8pt}{8pt}\selectfont Yes &  
         \fontsize{8pt}{8pt}\selectfont \makecell{Positively \\ correlated} & 
         \fontsize{8pt}{8pt}\selectfont Mixed&
         \fontsize{8pt}{8pt}\selectfont \makecell{
         Asymmetrical; \\Uniswap v3 AMM 
         } &
         \fontsize{8pt}{8pt}\selectfont \makecell{Perpetual \\Prorocol V2}\\ 
         \hline 
         
         \fontsize{8pt}{8pt}\selectfont \makecell{Oracle Pricing\\ Model} &   
         \fontsize{8pt}{8pt}\selectfont No &
         \fontsize{8pt}{8pt}\selectfont \makecell{Positively \\ correlated} &  
         \fontsize{8pt}{8pt}\selectfont \makecell{Negatively \\ correlated} & 
         \fontsize{8pt}{8pt}\selectfont \makecell{
         Asymmetrical;\\ Shalen’s dispersion \\of beliefs model \cite{shalen1993volume} 
         } &
         \fontsize{8pt}{8pt}\selectfont GMX; GNS\\ 
\hline
\end{tabularx}
\end{table}

\section{Acknowledgement}
ChatGPT has been used in the writing for the purpose of spelling, syntax and grammar checks. The authors take full responsibility of the output.

\ledgernotes


\newpage

\bibliographystyle{nature}

\newpage
\section{Appendix A}\label{Appendix.A}

In this section, we illustrate designs of the perpetual future contracts trading system for DEXs and CEXs, with emphasis on their differences and similarities. 

\subsection{Perpetual Futures}

Perpetual futures, introduced into the cryptocurrency market by BitMEX (\href{https://bitmex.com/}{https://bitmex.com/}) in 2016 and conceptualized by Robert J. Shiller\cite{shiller1993measuring}, represent a novel variant of traditional futures contracts. Unlike their traditional counterparts, perpetual futures do not have a fixed delivery date, allowing for indefinite trading without the need for contract rollovers. These contracts employ a funding fee mechanism to align their trading prices with the underlying spot prices, a process dependent on the price disparity between the perpetual futures and spot markets. This unique feature, along with the absence of a predetermined expiration date, distinguishes perpetual futures from traditional futures, where the price differential (basis) narrows as the expiration date approaches.
\begin{figure}
    \centering
    \includegraphics[width=0.6\linewidth]{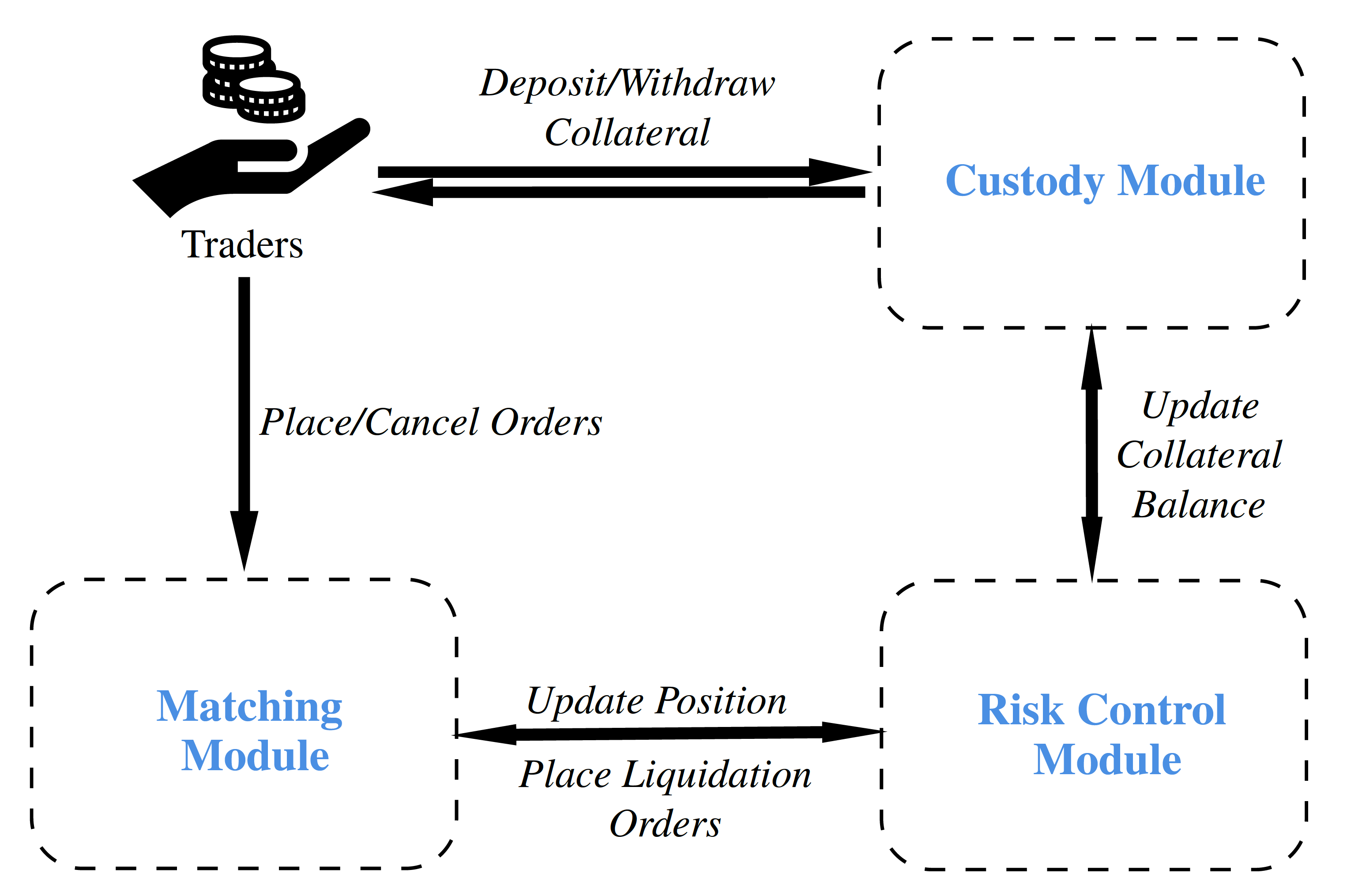}
    \caption{Elements of Exchange Systems}
    \label{fig:basic_term}
\end{figure}
Recently, perpetual futures have emerged as the most popular financial derivative in the cryptocurrency market, with trading volumes in 2021 and 2022 reaching \$51,989bn and \$39,806bn\cite{he2022fundamentals}, respectively, surpassing other derivatives like options\footnote{\href{https://www.theblock.co/data/crypto-markets/options}{https://www.theblock.co/data/crypto-markets/options}.}. Despite a significant portion of this trading occurring on centralized platforms, the rise of Decentralized Finance (DeFi) has enabled trading on decentralized platforms, with DEXs experiencing a faster growth rate in trading volume compared to CEXs. The main models for perpetual futures trading on DEXs include the Hybrid Model, Oracle Pricing Model, and Virtual Automated Market Making (VAMM) Model, each offering different approaches to managing and executing trades in the evolving landscape of digital asset trading.

\subsection{Models of Exchange Systems}\label{Sok_comparison}
The key stakeholders and elements are as follows:

\begin{table}
    \centering
    \caption{Comparison of different exchange models}
    \label{tab:features_comparison}
    \begin{tabularx}{0.9\textwidth}{ c  c  c  C c c}
    \hline 
         \tableheading{Type} &  
         \tableheading{Custody} &  
         \tableheading{Matching} &  
          \tableheading{Risk Control} & 
          \tableheading{Counterparty} &
          \tableheading{Pricing} \\ 
         \hline 
         
         CEX&  
         \makecell{Hot \& \\Cold Wallet} &  
         \makecell{Off-chain\\Server} &  
          \makecell{Off-chain \\Server} & Traders &Traders \\ 
         \hline 
         
         Hybrid &  
         \makecell{Smart \\contract} &  
         \makecell{Off-chain \\Server} & 
          \makecell{Off-chain \\Server} & Traders &Traders \\ 
         \hline 
         
         \makecell{Oracle \\Pricing} &  
         \makecell{Smart \\contract} &  
         \makecell{Smart \\contract} &  
          \makecell{Off-chain \\Keepers} & \makecell{Liquidity \\Providers} &Oracle \\ 
         \hline 
         
         VAMM &  
         \makecell{Smart \\contract} &  
         \makecell{Smart \\contract} &  
          \makecell{Off-chain \\Keepers} & \makecell{Liquidity \\Providers} &\makecell{AMM \\Algorithm} \\ 
         \hline
    \end{tabularx}
\end{table}
\begin{figure}
    \centering
    \includegraphics[width=0.7\linewidth]{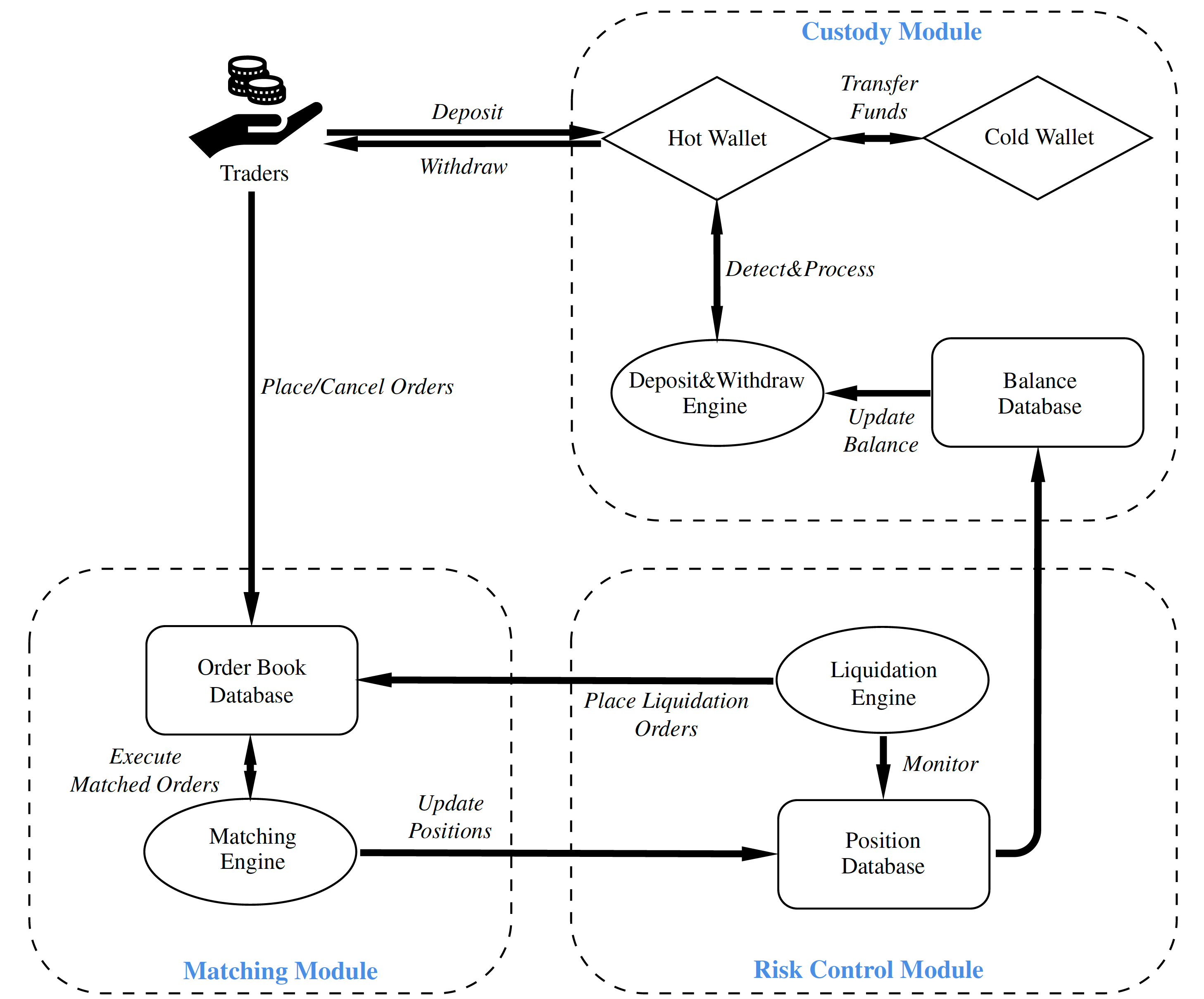}
    \caption{CEX Model. In this diagram, diamonds represent on-chain smart contracts, rectangles represent off-chain databases, and ellipses represent off-chain servers.}
    \label{fig:cex_model}
\end{figure}

\begin{figure}
    \centering
    \includegraphics[width=0.7\linewidth]{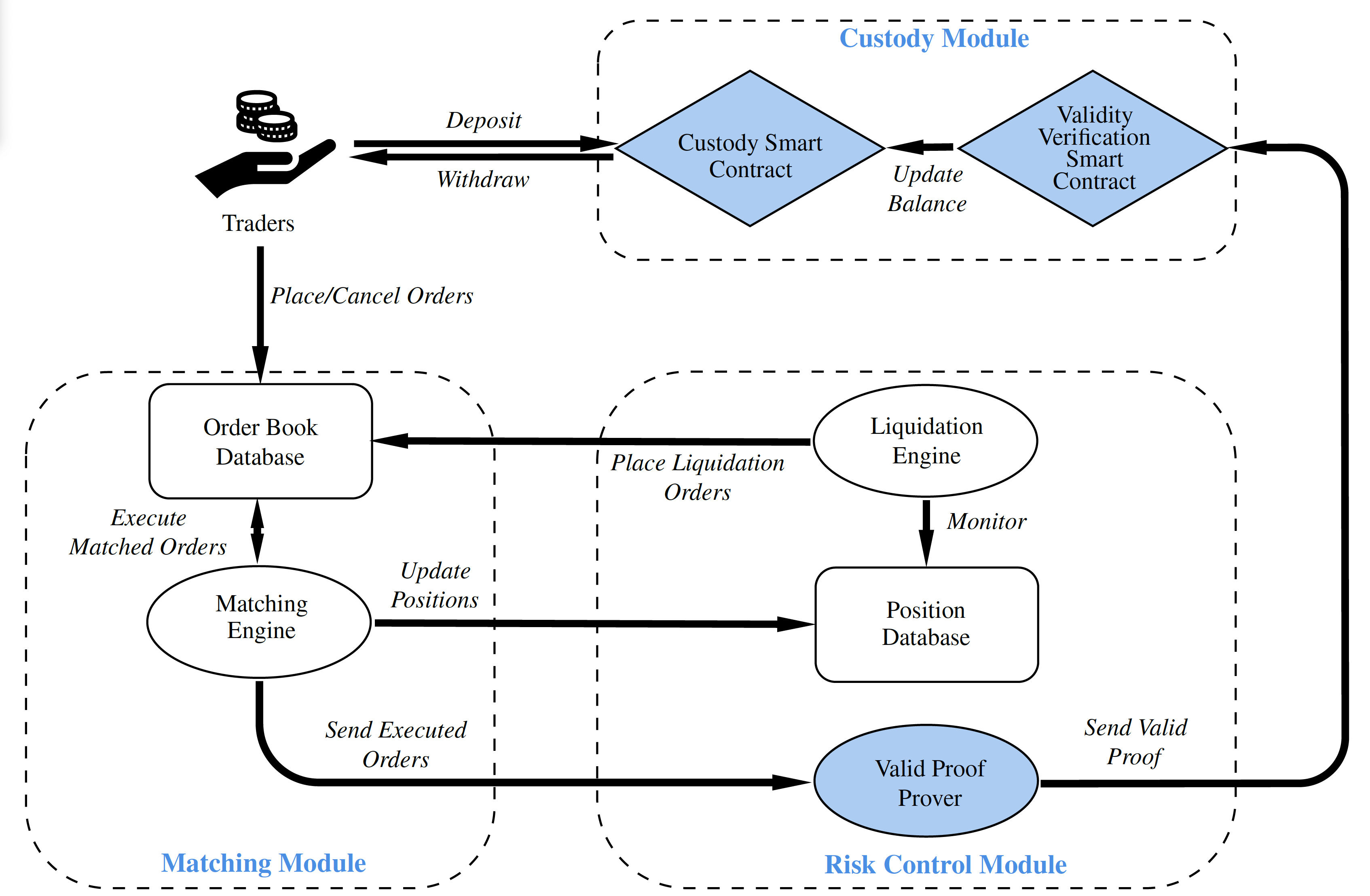}
    \caption{Hybrid Model. In this diagram, diamonds represent on-chain smart contracts, rectangles represent off-chain databases, and ellipses represent off-chain servers. We use blue background to indicate elements differentiating Hybrid Model from the CEX Model.}
    \label{fig:Hybrid_Model}
\end{figure}


\begin{figure}
    \centering
    \includegraphics[width=0.7\linewidth]{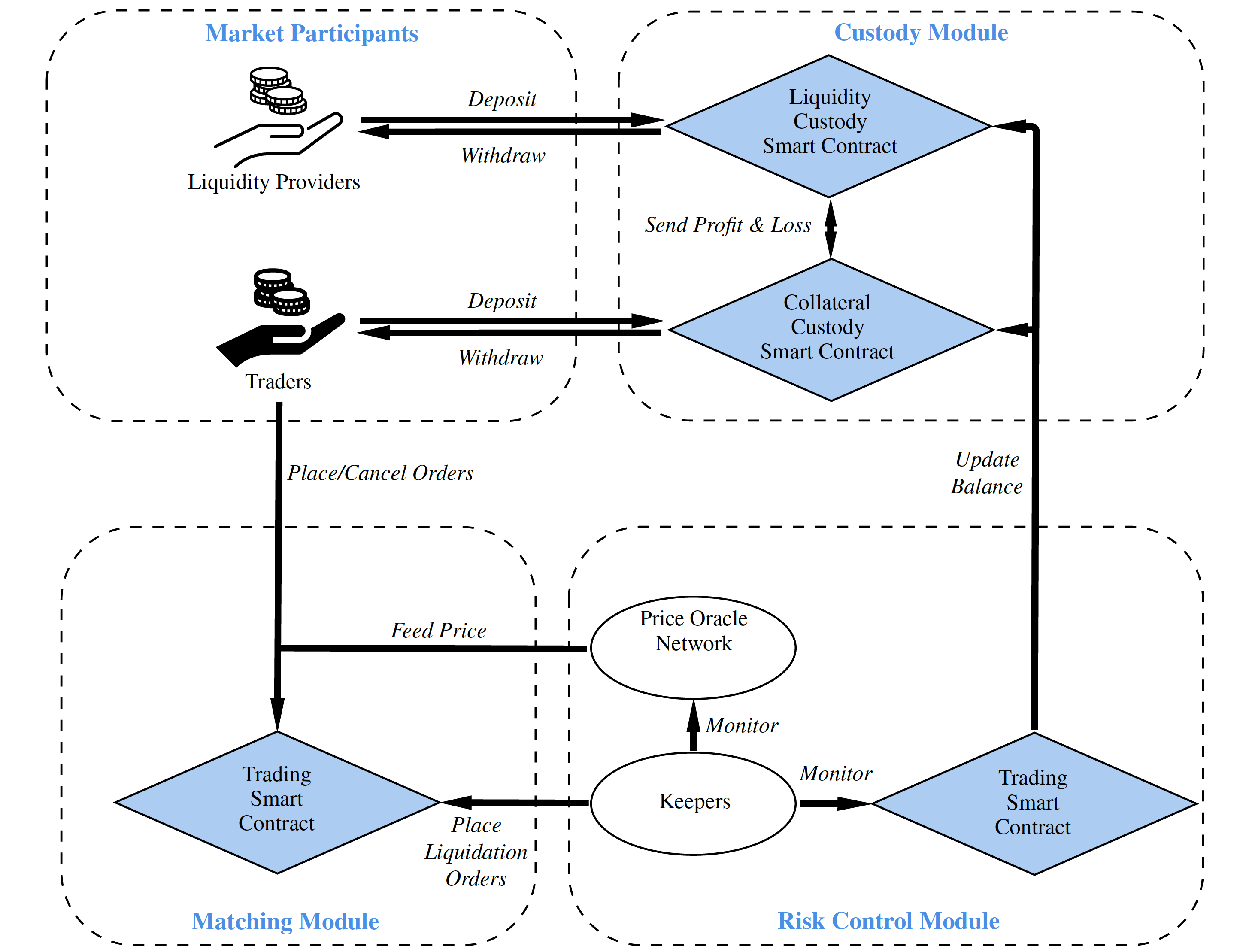}
    \caption{Oracle Pricing Model. In this diagram, diamonds represent on-chain smart contracts, rectangles represent off-chain databases, and ellipses represent off-chain servers. Blue background is used to indicate elements differentiating Oracle Pricing Model from the CEX Model.}
    \label{fig:Oracle_Pricing_Model}
\end{figure}


\begin{figure}
    \centering
    \includegraphics[width=0.7\linewidth]{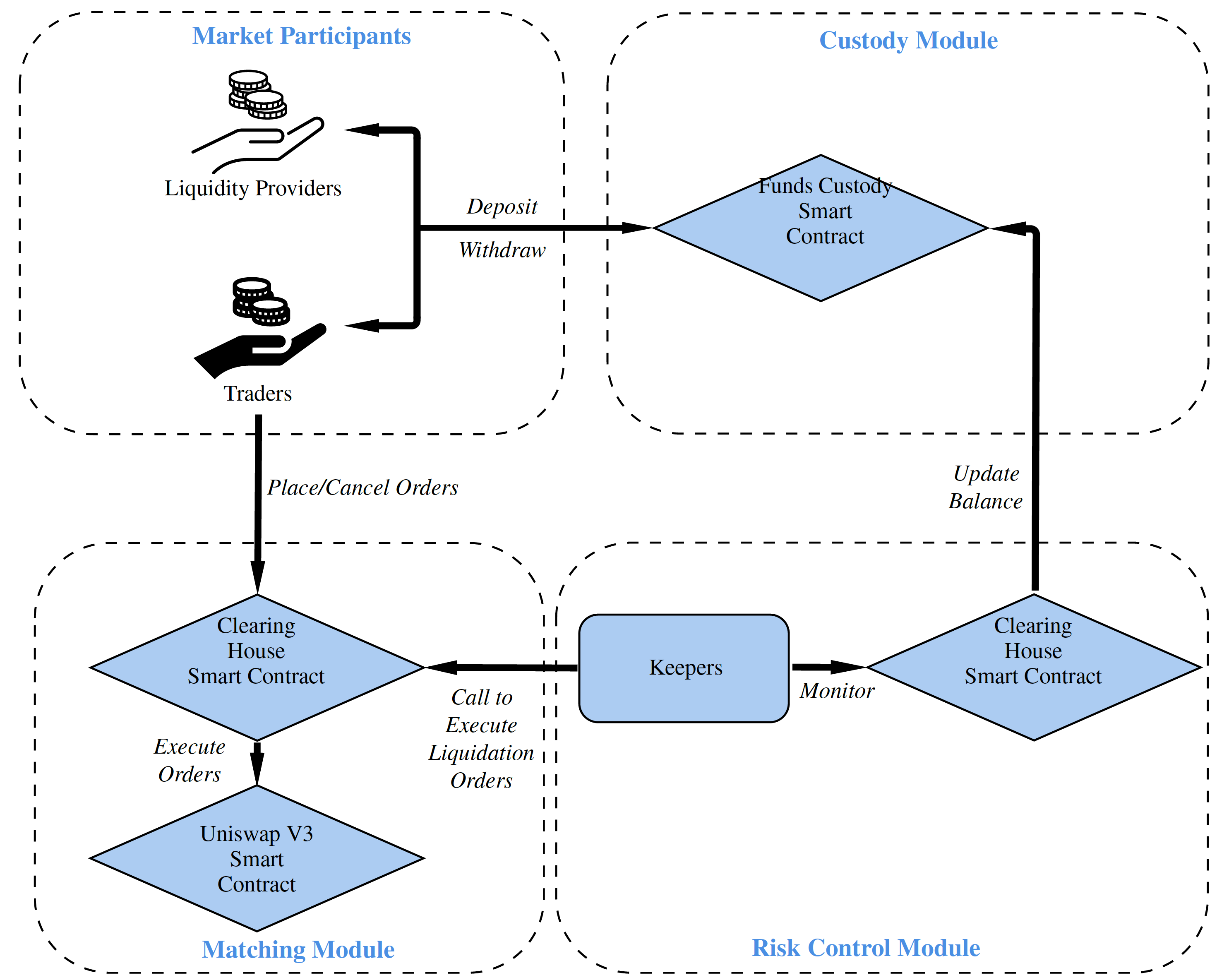}
    \caption{VAMM Model. In this diagram, diamonds represent on-chain smart contracts, rectangles represent off-chain databases, and ellipses represent off-chain servers. Blue background is used to indicate elements differentiating VAMM Model from the CEX Model.}
    \label{fig:Virtual_Auto_Market_Making_Model}
\end{figure}


\begin{itemize}
    \item Trader: Individuals or entities engaged in the purchase and sale of perpetual contracts. These traders furnish collateral to maintain and manage their positions through the trading of such contracts.
    \item Custody Module: This module is responsible for ensuring the security of assets across all trader accounts. It consistently updates and retains the latest balance details and facilitates both deposit and withdrawal operations initiated by traders.
    \item Matching Module: This module is entrusted with storing, correlating, and executing purchase and sale orders of contracts.
    \item Risk Control Module: This module is vital for assessing and supervising the position of every trader account, contingent on the orders that have been executed. Its role is pivotal in ascertaining that the provided collateral is sufficient to offset potential deficits. Furthermore, in specific scenarios, it assumes control of a trader's position and proceeds with its liquidation.
\end{itemize}
How those four parts interact is illustrated in Fig.~\ref{fig:basic_term}.

The modules employed by different trading systems (models) vary significantly. In Table~\ref{tab:features_comparison}, we provide a horizontal comparison across five dimensions: Custody, Matching, Risk Control, Counterparty, and Pricing.

The Custody Module assumes a central role in demarcating the distinct classifications within the realm of perpetual futures exchanges. Those that employ centralized custody solutions are classified as Centralized Exchanges (CEX Model) (Fig.~\ref{fig:cex_model}), whereas those that abstain from such practices fall under the rubric of Decentralized Exchanges (DEX). Within the domain of DEX, further subdivisions emerge contingent upon factors encompassing order matching, price determination, and counterpart matching, resulting in  to the Hybrid Model (Fig.~\ref{fig:Hybrid_Model}), Oracle Pricing Model (Fig.~\ref{fig:Oracle_Pricing_Model}), and Virtual Auto Market Making (VAMM) Model (Fig.~\ref{fig:Virtual_Auto_Market_Making_Model}). Since both CEX Model and Hybrid Model adopt limit-order-book, these two categories of exchanges could also be regarded as Limit-Order-Book (LOB) Model.

The CEX Model(Fig.~\ref{fig:cex_model}) closely resembles traditional financial practices. Apart from deposits and withdrawals, it makes limited use of blockchain technology. The technology stack used for custody of client funds, order matching, settlement, risk assessment, and control closely aligns with traditional financial practices. Due to the imperfect state of infrastructure and regulatory rules in the cryptocurrency trading industry, CEX Model often combines the functions of custody banks, traditional exchanges, clearinghouses, and brokerages entities.

In contrast, the Hybrid Model(Fig.~\ref{fig:Hybrid_Model}) harnesses smart contracts for custody and settlement processes, ameliorating the centralization quandaries that typify the CEX Model while retaining certain centralized elements. Notably, off-chain servers persist in facilitating order matching and counterpart matching functions.

The Oracle Pricing Model(Fig.~\ref{fig:Oracle_Pricing_Model}) and VAMM Model (Fig.~\ref{fig:Virtual_Auto_Market_Making_Model}) gravitate toward a greater degree of decentralization. Both models leverage smart contracts for order matching, with Liquidity Providers assuming the role of direct counterparties to traders, engendering an indirect mode of trade execution among traders. Nevertheless, it is imperative to acknowledge that both models still rely upon centralized constituents. The involvement of Oracles and Keepers in the Risk Control Module remains a requisite for effecting the liquidation process. Moreover, the Oracle Pricing Model hinges upon Oracles for the determination of precise trade prices.

The augmentation of decentralization engenders certain trade-offs. While diminishing reliance on centralized components can mitigate specific security vulnerabilities and fortify resistance against censorship, it concurrently imposes constraints on the exchange's transaction processing capacity per second (TPS). As one progresses from the CEX Model to the Hybrid Model, Oracle Pricing Model, and VAMM Model, the degrees of centralization diminish. However, correspondingly, the ability to handle transactions experiences a commensurate decline due to the TPS limitations inherent in the underlying blockchain network. Consequently, blockchain developers have embarked upon the development of diverse technologies aimed at enhancing throughput and efficiency. Examples include zero-knowledge scaling solutions for Ethereum (zkEVM), which offer reduced costs, faster transaction speeds, and scalable solutions, all while maintaining the core principles of security and decentralization.

\section{Appendix B}\label{Appendix.B}

\begin{table}[b]
\caption{Granger Causality Test of Return on Bitcoin and the Net Change in Liquidity in the vBTC-vUSDC Pool on Perpetual Protocol V2}\label{tab_grangerCausality}

\begin{tabularx}{\textwidth}{ c  c  c  C }

\hline 
 \tableheading{$H_{0}$}\footnote{This table reports the results of Granger Causality Test in dectacting the relationship between return on Bitcoin and the net change in the liquidity contained in vBTC-vUSDC pool on Perpetual Protocol V2. }& 
 \tableheading{Max-lag}& 
 \tableheading{F-statistics}&
 \tableheading{p-value\\($Prob>F$)}\\
 
\hline
\makecell{Return on BTC does not Granger-cause\\ net change in liquidity pool} & 
15 & 
1.624 & 
0.079 \\

\hline
\makecell{Net change in liquidity pool\\does not Granger-cause Return on BTC} & 
15 & 
1.164 & 
0.311 \\

\hline

\end{tabularx}

\end{table}

As conjectured in our analysis on liquidation in Perpetual Protocol V2, we hypothesis that price rise in Bitcoin attract liquidity providers in offering liquidity. To detect empirical evidence to our conjecture, we conduct the Granger Causality Test to see the relationship between the return on Bitcoin and the net change in liquidity in the vBTC-vUSDC pool on Perpetual Protocol V2, where the net change is calculated by subtracting the daily added liquidity by the daily withdrawn liquidity from the pool (measured in Bitcoin). Return on Bitcoin is calculated in daily frequency with the log-form return as follows: $R_{t}=\log \left(\underline{P_t}\right)-\log \left(\overline{P_t}\right)$, where $\underline{P_t}$ and $\overline{P_t}$ denote the close and open price of Bitcoin on day $t$. We retrieve data from 15 May 2023 to 15 Oct 2023 on Etherscan (\href{https://optimistic.etherscan.io/}{https://optimistic.etherscan.io/}). The testing results are reported in Table~\ref{tab_grangerCausality}. There exists Granger-causality from the return on Bitcoin to the net change in vBTC-vUSDC liquidity pool, indicated by the statistically significant F-statistic with p-value of 0.079. However, the causality in reversed direction (net change in vBTC-vUSDC liquidity pool Granger-cause return on Bitcoin) is not evidenced, with p-value of 0.311, supporting our hypothesis that more liquidity is injected when Bitcoin price jump. The large lag, i.e., 15 days in this case, may suggests that traders need time to adjust positions in the liquidity pool in reaction to the rise in Bitcoin price. The empirical evidence here shows a rough picture of how traders change their deposit in the liquidity pool while future studies are needed with more factors controlled.

\thispagestyle{pagelast}

\end{document}